# HYPERVISOR-BASED ACTIVE DATA PROTECTION FOR INTEGRITY AND CONFIDENTIALITY OF DYNAMICALLY ALLOCATED MEMORY IN WINDOWS KERNEL

Igor Korkin, PhD
Security Researcher
Moscow, Russia
igor.korkin@gmail.com

## ABSTRACT

One of the main issues in the OS security is providing trusted code execution in an untrusted environment. During executing, kernel-mode drivers dynamically allocate memory to store and process their data: Windows core kernel structures, users' private information, and sensitive data of third-party drivers. All this data can be tampered with by kernel-mode malware. Attacks on Windows-based computers can cause not just hiding a malware driver, process privilege escalation, and stealing private data but also failures of industrial CNC machines. Windows built-in security and existing approaches do not provide the integrity and confidentiality of the allocated memory of third-party drivers. The proposed hypervisor-based system (AllMemPro) protects allocated data from being modified or stolen. AllMemPro prevents access to even 1 byte of allocated data, adapts for newly allocated memory in real time, and protects the driver without its source code. AllMemPro works well on newest Windows 10 1709 x64.

**Keywords**: hypervisor-based protection, Windows kernel, Intel, CNC security, rootkits, dynamic data protection.

## 1. INTRODUCTION

Currently, protection of data in computer memory is becoming essential. Growing integration of ubiquitous Windows-based computers into industrial automation makes this security issue critically important. Windows machines can be attacked when malware kernel-mode code manipulates the memory content of legal drivers and their dynamically allocated memory pools, which store critical data.

Intruders can tamper with this data by installing their own malware driver or using vulnerabilities of the installed kernel mode modules (Adler, 2017).

There are a number of vulnerabilities in Windows kernel core drivers as well as in the third-party drivers such as NVIDIA Windows GPU Display Driver (NVIDIA Corporation, 2017), Audio Driver (Gee, 2017), keyboard driver (CSO, 2017), Schneider Electric UnitelWay Device Driver (Langill, 2011). For example, an attacker could successfully exploit the CVE-2017-0155 vulnerability in the Win32k component and run arbitrary code in kernel mode (Microsoft, 2017).

The vulnerable VirtualBox driver (VBoxDrv.sys) has been exploited by Turla rootkit and allows to write arbitrary values to any kernel memory (Singh, 2015; Kirda, 2015).

Another vulnerability of CPU-Z driver has been exploited in HandleMaster project change granted access rights for handles (MarkHC, 2017).

Additionally, in a recent paper 'Windows exploitation in 2016' researchers from ESET underline the vulnerability of third-party drivers as a real vector of exploitation (Baranov, 2016).

This malicious code is running at the same privilege level as a Windows kernel. There are no built-in Windows security control features to prevent illegal malware access in the kernel mode.

As a result, intruders can tamper with the following allocated data in the kernel-mode, see Figure 1:

- Windows core kernel;
- User data;
- Industrial automation control software.

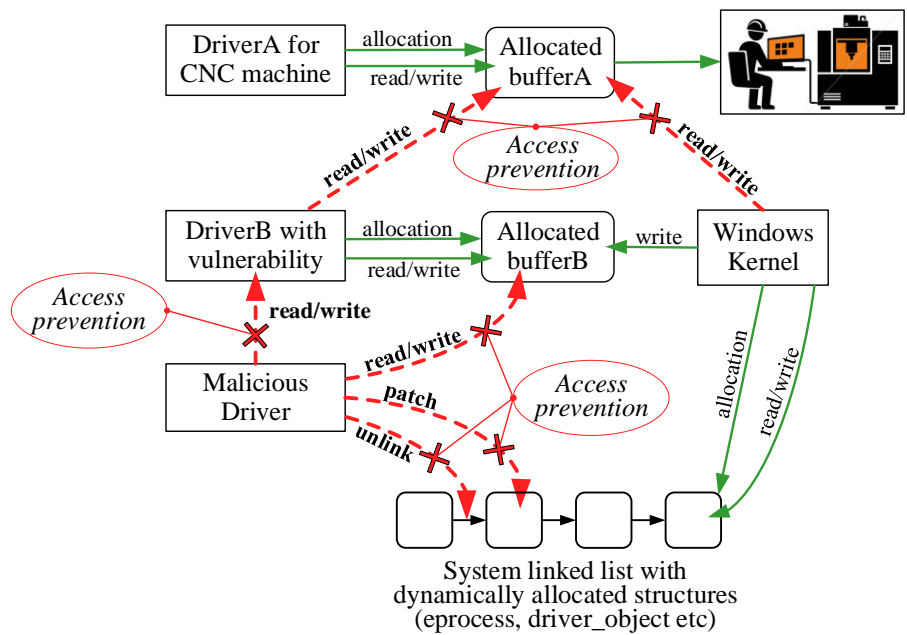

Figure 1 Examples of driver's memory access attempts to the allocated memory:
legitimate access attempts are in green, unauthorized ones are dashed red arrows

**Windows kernel security issues**. Firstly, hackers patch allocated system data in Windows kernel to prevent the detection of installed malware drivers and escalate process privileges. Information about a loaded driver is collected in several system lists, which include allocated structures connected by linked lists. Hackers can unlink the structure of malware driver from all these lists to make them hidden. Consequently, bytes in these structures in memory can be deliberately changed and made useless in finding malware footprints. These rootkit techniques are known as DKOM. Also, rootkits can read the undocumented kernel-mode values. For example, DisPG disables Windows Kernel Patch Protection by using the undocumented value nt!PoolBigPageTableSize, which needs to be protected (Korkin & Tanda, 2017).

Usually, malware processes are running with low privilege and obtaining high privilege can allow the malware to perform more operations. Process privileges can be escalated by exploiting kernel mode driver bugs using SID List Patching, Privileges Patching, and Token Stealing payloads (Perla & Oldani, 2010; Hasherezade, 2017). The issue of finding and exploiting kernel-mode vulnerabilities is quite challenging and powerful because it could allow compromising the system completely (Cisco, 2017).

Another example of escalating process privileges is the mimikatz framework, which loads its own driver and manipulates Token value from EPROCESS structure (Delpy, 2018).

The ProjectSauron is one of the examples of kernel-mode malware drivers, which are classified as Advanced Persistent Threat. It supports commands to elevate privileges to a system account (Kaspersky, 2016).

One more example is the CVE-2016-7255 exploit, which uses type-confusion vulnerability in win32k.sys (CVE-2016-7255) to gain elevated privileges by patching EPROCESS structure (Oh & Florio, 2017). The similar Elevation-of-Privilege (EOP) attack was used in Duqu 2.0 exploit (Wook & Florio, 2015).

Another exploit overwrites the Server Message Block (SMB) connection session structures to gain Admin/SYSTEM session (Rapid7, 2018).

Apart from EPROCESS structures attackers are also patching other allocated Windows objects.

TDL bootkit conceals its presence in the system by modifying the StartIo field of the target device's driver and excluding the target device from the DRIVER_OBJECT's linked list (Rodionov & Matrosov, 2011).

Win32/Gapz and ProxyBox hook IRP handlers from DRIVER_OBJECT.MajorFunction[] array to protect itself from being removed from the system. (Rodionov & Matrosov, 2013; Bingham, 2012).

Another technique is to manipulate IRP structure and IO_STACK_LOCATION. For example, by changing CompletionRoutine pointer it is possible to avoid the calling of completion routine of a filter driver, which prevents collecting evidence of suspicious activity (Blunden, 2009).

Windows 10 supports Supervisor Mode Execution Prevention (SMEP), which prevents the kernel from executing code in user mode, a common technique used for local kernel elevation of privilege (EOP). SMEP requires processor support found in Intel Ivy Bridge or later processors, and it also can be bypassed (Shahat, 2018).

The protection system needs to provide integrity for the sensitive memory objects, linked lists as well as preventing modification of each data structure in these lists, which has been dynamically allocated by Windows kernel. Additionally, the security software needs to prevent illegal reading of critical Windows kernel values.

**User data issues**. Secondly, malware can attack the user data located in the kernel memory. For example, malware could steal or overwrite private keys, which are used in cryptographic drivers and can be used to decrypt user data. Malware can also attack a user's privacy by reading Windows telemetry and other data collected by Windows OS.

The protection system needs to prevent unauthorized access to memory data dynamically allocated by third-party drivers.

**Security of industrial automation control software**. Thirdly, malware can cause considerable damage by attacking industrial control software. Stuxnet is a famous example of kernel-mode malware, which took out 1000 centrifuges at Iran's nuclear facility. Its driver deliberately attacked Windows-based Industrial Control Systems (ICS), which is used with Siemens PLC (Anderson et al., 2017).

A similar vector for cyber-attacks is the software for the machines with computer numerical control (CNC), which use computers to handle various machine tools: lathes, grinders, drilling machines etc. CNC machines are extremely popular in the cutting-edge manufacturing, for example, they are used by NASA, Boeing, SpaceX etc. (Isakovic, 2018).

Cybersecurity researchers emphasize a substantial risk of cyber-attacks against manufacturing systems and CNC machines (Vincent et al., 2015; Chhetri et al., 2016). Researchers illustrate various threatening scenarios related to remote maintenance of CNC machines (Mehnen et al., 2017). Security experts from the USA stress the serious security risks for CNC machines and provide nine basic protection tasks to do before connecting machine tools to the network. One is to "install a Windows anti-virus protection service" (Johnson, 2017). However, there are no specific AV services to protect industrial CNC machines; and it is obvious that basic desktop AVs cannot provide full protection for CNC machines.

The recent report from a German IT association says that more than a half of companies in Germany "have been victims of industrial espionage, sabotage or data theft in the last two years" and these "companies had incurred a loss of around 55 billion euros per year", which is around 64 billion USD (Burgess, 2017).

Another report released by Trend Micro Threat Research and Polytechnic University of Milan demonstrates the cyber-attack implementations on actual industrial robots. (Maggi et al., 2017)

There are several Windows-based CNC control systems such as Fanuc (Fanuc, 2017), MACH series (Mach, 2018), UCCNC (CNCdrive, 2018), MicroSystems (WinCNC, 2018), which use kernel-mode drivers and dynamically allocate data in kernel mode to send the control codes to the machines. For example, UCCNC software provides an opportunity to control professional CNC machines using a tablet PC running on Windows 8.1. (Stoney CNC, 2015; Bozso777, 2015). Another example is Fanuc software, which leverages a WinIO library and a parallel port to manage CNC machines (DSP, 2011).

All this industry control software can be potentially attacked by a malware driver, which can accomplish both types of attacks: sabotage and industrial espionage. The first type of attacks can result in crashing the machine or "workpiece damage such as spalling or delamination" (Schulze et al., 2011). The second type of attacks can result in sensitive information being stolen to reconstruct the workpiece or the technologies of processing, such as know-how, trade secrets, and confidential information.

The Deutsche Welle reports that one of the CNC machines in a German engineering firm was attacked by spyware: "it turns out their computer controlled molding cutter came spiked with sophisticated malware that automatically transferred sensitive data about the new prototype to Asian-based Internet Protocol Addresses" (Knigge, 2013). The CNC machines security is a sensitive topic and existing hush-hush culture makes it very difficult to come up with concrete examples of the incidents of cyber-attacks.

The protection software needs to prevent illegitimate access to dynamically allocated buffers in kernel-mode used by industry software drivers to control CNC machines.

**Threat model**. Summing up all malware actions, I would like to introduce a threat model. The following malware activities will be considered in this paper, see Figure 1:

- Intruders avoid all prevention measures and can install kernel-mode malware;
- Malware driver easily finds the memory content with sensitive data and code;
- Malware driver reads and writes the memory data allocated by Windows kernel and any third-party drivers;
- Malware reads and writes code sections of kernel modules.

The Windows operating system includes several features to prevent illegal memory access. However, these features provide only the integrity of code sections of kernel modules and check the integrity of systems linked lists. They do not provide the integrity of each structure from these lists and do not prevent reading of any kernel memory.

There are several research projects, which partially solve the problem with protection of allocated data in the kernel.

In the paper *Kernel Data Integrity Protection via Memory Access Control* by Srivastava et al., 2009 the authors proposed using a hypervisor to mediate the execution of instructions attempting to write protected kernel data. Their system prevents overwriting only Windows OS critical data: process credentials for privilege escalation and detects illegal removing structures from linked lists. This system does not prevent confidentiality breach of kernel data and code nor does it track allocation of memory pools to protect them.

In the paper *HACS: A Hypervisor-Based Access Control Strategy to Protect Security-Critical Kernel Data* by Wang et al., (2017) the authors focus on rootkits that place a malicious code in their installation procedure. HACS maintains a module whitelist: only the modules in the whitelist are legal to modify the protected region. Consequently, HACS does not have an opportunity to provide various memory access policies for various kernel modules. Given this opportunity, HACS could prevent illegal memory access if one of the trusted drivers is compromised.

Another system DADE (Yi et al., 2017) provides kernel integrity via periodically scanning invariant properties and checking backtraces of kernel function calls. The invariants describe the behavior expected from an uncompromised kernel using the source code of OS kernel. Authors admit that DADE has only a probabilistic chance to detect integrity attacks. Neither does DADE protect the memory of third-party drivers.

As a result, the issue of providing the integrity and confidentiality of dynamically allocated data in the kernel mode is unsolved. At the same time, the illegal memory access to this data can result in not only hiding malware drivers and stealing users' private data, but also in damaging industrial process and stealing know-how.

The goal of this paper is to tackle this issue. First, I propose an memory access rules to deal with kernel-mode malware, which have the following main principles:

- Trap each memory access and grant full access only to the memory data, which has been allocated by this driver before;
- Prevent unauthorized access even to one byte of memory allocated by another driver;
- Provide integrity and/or confidentiality of the allocated data according to the principle of least privilege;
- Recover memory content after it was modified.

These memory access rules have to be adapted in real time to changing situations in the OS and help to restore the original data:

- Separate memory access rules for each kernel-mode driver;
- Automatically update a list of memory access rules when a new driver is loaded or any module calls allocation or deallocation routines.

The purpose of this paper is to present the design, implementation, and evaluation of a new hypervisor-based system that protects the dynamically allocated data in Windows kernel from being accessed without authorization.

The remainder of the paper proceeds is as follows:

Section 2 provides the review of newest kernel mode memory protection features, which have been integrated into Windows 10 1709. The comparative analysis of the existing memory protection approaches will be given according to the proposed threat model.

Section 3 contains the architecture of proposed allocated memory protection system – AllMemPro and the experimental results.

Section 4 focuses on the main conclusions and further research directions.

## 2. BACKGROUND

This chapter presents the analysis of the existing papers and software tools that are focused on allocated memory protection in OS kernel. Popular research projects are compared by their capability to provide integrity and confidentiality to dynamically allocated memory and code in kernel-mode for OS core kernel and third-party drivers.

There are several new protection facilities which have been integrated on Windows 10 1709 and have significantly improved the security of the OS: Device Guard, Credential Guard, UEFI Secure Boot, updated Kernel Patch Protection (PatchGuard), Supervisor Mode Execution Prevention (SMEP), Early Launch Antimalware (ELAM), Windows Defender Exploit Guard (WDEG) etc. Device Guard and PatchGuard are dealing with the issues of memory protection in kernel-mode memory (Zylva, 2016; Hall et al., 2017).

Device Guard includes three basic components, one of them being the Kernel Mode Code Integrity (KMCI) which prevents patching executable pages in the kernel memory. According to the "Driver compatibility with Device Guard in Windows 10" memory pages and sections can never be writable and executable simultaneously and executable code cannot be directly modified (Baxter, 2017).

PatchGuard protects critical structures in the Windows kernel from modification by unknown code. It stores and periodically verifies checksums of specific kernel memory areas. PatchGuard causes a BSOD if a mismatch is found. "Kernel patching can result in unpredictable behavior, system instability, and performance problems like the Blue Screen of Death" (Field, 2006). For linked lists, PatchGuard checks only the integrity of the links between structures and has 4 various types of BSOD (Marshall, 2017), without preventing the structure modification. In addition, PatchGuard reveals the corruption of MajorFunction table in DRIVER_OBJECT, but whether it protects other fields or not is undocumented (Mei, 2014; OSR, 2016).

In summary, Windows built-in security features provide the integrity of the following:

- code sections of kernel-mode modules;
- undocumented internal lists with allocated structures.

Windows security features do not support the integrity and confidentiality of allocated memory of third-party drivers. In addition, Windows PatchGuard does not prevent illegal memory modifications; it just causes a BSOD in case of them. These BSODs are not appropriate for use in critical infrastructure, like CNC-machines and industry.

Security experts investigate these security issues and propose various solutions to protect sensitive data in the kernel mode. All projects can be divided into two groups that are based on kernel-mode drivers and hardware virtualization. The scope of this paper is the hypervisor-based solutions because they work in a more privileged mode than kernel-mode malware and are resilient to its attacks.

All hypervisor-based approaches can be divided into two subgroups according to the technologies which they use to intercept memory access in the kernel-mode (Korkin & Tanda, 2017).

The first subgroup controls memory access to the sensitive data by marking a memory page with this data as non-present. Next, each access to this page generates a page-fault exception (#PF), which will be trapped and dispatched by the hypervisor.

The second subgroup leverages a new Intel VT-x with Extended Page Tables (EPT) technology. EPT mechanism can separately intercept read-, write-, and execute access. The hypervisor allows or disallows access to the memory page by setting bits in the EPT memory structures. Thus, any disallowed access will involve EPT violation and will be processed by the hypervisor.

EPT mechanism is faster than #PF-based one. EPT-based approach can intercept, for example, only write- access and skip others, while #PF-based one always intercepts all access to the memory page. At the same time, #PF-based approaches are working on all computers, while EPT has been integrated into the Intel CPUs since Nehalem microarchitecture.

Srivastava et al., (2009) confirm that "kernel-level malicious software has full access to the data and operations of all kernel components." Their protection system is based on page-fault exceptions and protects kernel variable and system data structure elements from being patched by malware. Their system Sentry provides only integrity of dynamically allocated data by partitioning kernel memory into two parts: protected and unprotected regions. The authors assume that the core kernel has full trust, while other drivers hold only limited trust. Sentry mediates the execution of instructions attempting to write protected kernel data and verifies memory access at the granularity of high-level language variables in the kernel's source code. Sentry has been developed using Linux and Xen hypervisor.

Thus, Sentry does not provide the following:

- flexible memory access policy to protected new allocated data by third-party drivers;
- confidentiality of data;
- kernel-mode code integrity.

Additionally, Sentry requires the kernel's source code, which is not applicable for Windows OS.

Another system, HACS (Hypervisor-Based Access Control Strategy) by Wang et al., (2017), leverages EPT technology to intercept write requests to the protected regions. HACS can detect modifications of security-critical kernel data and escalation process privileges by setting read-only access rights to the corresponding memory pages. Authors proposed to use a whitelist-based access control strategy. The whitelist, which is made by user experience, contains only credible kernel modules. One of HACS's features is detecting memory modifications from a malware code located in the initialization procedures. This system is implemented on BitVisor version 1.4 and tested on Ubuntu version 14. This project includes the same disadvantages as the previous one and HACS provides just two levels of trust: legitimate or illegitimate modules. As a result, this system cannot prevent access from a legitimate module to the memory area, allocated by another legitimate one.

The project DADE (Data Anomaly Detection Engine) by Yi et al., (2017) performs memory introspection and verifies the integrity of kernel data by checking whether certain integrity specifications hold or not. Authors propose to use EPT facilities to intercept write access. DADE marks memory pages with protected data as read-only, and then any write access to this page generates a page fault, which is handled by the hypervisor. The key idea of DADE design is to leverage the information available at object allocation events, namely backtraces of kernel function calls. For example, a malware module attempts to remove their structure from a system list and produces a specific deallocation event backtrace. DADE compares this backtrace with a legitimate one, which is produced by core kernel when a module is unloaded. It is obvious that these backtraces are different and DADE reveals the unlinking attack. The DADE prototype has been implemented using KVM hypervisor with Linux version 3.8.0. DADE requires a source code of OS kernel.

The issues of preventing commodity OS kernel from vulnerable loadable kernel modules are analyzed in the project LKMG (Loadable Kernel Module Guard) by Tian et al. (2018), which is related to the second subgroup and uses EPT technology. LKMG can reveal the following malware activities in the kernel: modification of code and data, calling unauthorized kernel functions and stealing kernel sensitive information. The authors propose to use a policy-centric system to isolate various loadable drivers from the rest of the kernel. These security policies are generated automatically from source codes. LKMG utilizes the general security policy for dynamic data access: a driver can only access its own allocated kernel mode regions. LKMG is based on the Xen hypervisor and protects only Linux OS.

In comparison with Sentry, HACS, and DADE, LKMG provides integrity and confidentiality for allocated data as well as code integrity. However, all of these require a source code of OS kernel, which is impossible for Windows OS.

Hypervisor-based system HUKO (Liu et al., 2011) protects the kernel integrity for commodity OS from untrusted extensions. It uses VT-x and EPT technologies and is able to track dynamic contents such as dynamic kernel data, stack and heap region, and loadable extensions. It is able to protect the integrity of both kernel code and data. HUKO prevents the OS data from being modified by kernel-mode drivers by isolating untrusted extensions from the OS kernel. HUKO considers three different categories of memory access subjects: OS kernel, trusted extensions, and untrusted extensions. HUKO does not restrict the OS kernel. Authors admit "it is possible that attackers can exploit the legitimate kernel interface to subvert the integrity of kernel," for example, by exploiting bugs of the kernel API functions. In addition, HUKO does not protect privacy and integrity of the kernel-mode data of third-party drivers from being tampered.

There are several research projects InkTag by Hofmann et al. (2013), AppGuard by Zha et al. (2015), which apply EPT technology to guarantee data security. However, they protect user-mode application contexts with code and data from the OS kernel and other apps and do not guarantee the security of kernel-mode memory.

The following research prototypes ExOShim by Brookes et al. (2016), HyperForce by Gadaleta et al. (2012), and Sprobes by Ge et al. (2014) prevent memory disclosure attacks and provide kernel-mode integrity, without data protection. These projects only partially solve the goals of this paper.

Authors proposed to apply the hypervisor-based system to reveal new DKOM attacks, which tamper with dynamic data structures. They considered the scenarios when malware subverts the OS scheduler and proposed an idea of detecting these anomalies by monitoring and checking the execution time of all processes. Their solution can only detect any unauthorized data modifications, without preventing or repairing them (Graziano et.el., 2016).

Security researcher A. Zabrocki proposed an advanced analog of PatchGuard for Linux-based OSes. Named Linux Kernel Runtime Guard (LKRG), it is a loadable kernel module that performs runtime integrity checking of the Linux kernel. LKRG supports from being loaded at early boot stage and "protects the system by comparing hashes which are calculated from the most important kernel region/ sections/structures with the internal database hashes". A current version of LKMG provides code integrity and exploit detection. But, it does not protect allocated memory of third-party drivers (Zabrocki, 2018).

The summary table with the comparative analysis of the major papers and projects is given in Table 1.

In addition, the vast majority of analyzed methods require the driver's source code to protect allocated data. The proposed AllMemPro system can protect the compiled code without its source code.

The next section will present the proposed system, which is said to be free from all the drawbacks mentioned above.

Table 1 Summary table of memory protection projects

| Title, year | OS Kernel | | Third-Party Kernel-Mode Drivers | | | | OS |
|---|---|---|---|---|---|---|---|
| | Integrity | | Integrity | | Confidentiality | | |
| | Allocated Data | Code | Allocated Data | Code | Allocated Data | Code | |
| Device Guard and Patch Guard in Windows 10 1709, 2017 | +—[A] | + | – | + | – | – | Windows |
| Sentry, 2009 | + | – | + | – | – | – | Linux |
| HUKO, 2011 | + | + | +—[B] | – | – | – | Windows Linux |
| HyperForce, 2012; Sprobes, 2014; ExOShim, 2016 | – | + | – | – | – | – | Linux |
| HACS, 2017 | + | – | – | – | – | – | Linux |
| DADE, 2017 | + | – | + | – | – | – | Linux |
| LKMG, 2018 | + | + | +—[B] | + | +—[B] | – | Linux |
| LKRG, 2018 | + | + | – | + | – | – | Linux |
| *AllMemPro, 2018* | + | –[C] | + | –[C] | + | –[C] | Windows |

[A] Windows security features reveal only unlinking critical structures; but they do not prevent changing the content of these structures;

[B] HUKO and LKMG systems do not restrict the OS kernel, and as a result, they only partially protect data, which have been allocated by third party drivers;

[C] The current version of AllMemPro protects only allocated data in the kernel mode. The protection of code integrity and confidentiality will be implemented further.

## 3. PROPOSED ALLOCATED MEMORY PROTECTION – ALLMEMPRO

This section covers the details of the proposed hypervisor-based system to guarantee the confidentiality and integrity of dynamically allocated data.

To start with, I will show how to apply a hypervisor and EPT technologies to prevent these three scenarios of attacks on kernel-mode memory. I am going to use memory access rules, which are actively adapted to the newly allocated/freed data.

Afterward, I will present the architecture of AllMemPro, which realizes the proposed ideas and will give some details about how to prevent unauthorized access to the allocated memory and grant access to the legitimate kernel-mode module.

Finally, I will show three cases of using the developed proof of concept prototype to protect allocated memory for both third-party driver and Windows kernel.

### 3.1. Apply EPT to Guarantee Integrity and Confidentiality of Dynamically Allocated Data

This section suggests using EPT technology to prevent typical malware attacks.

As was stated above, the dynamically allocated data can contain sensitive information, such as crypto keys, users' private data, parameters of CNC machines, process privileges and drivers information.

**Three Scenarios of Attacks**. Attackers try to tamper all this data and it is possible to define three main scenarios, see Figure 1. First, attackers can steal/read and modify/write the allocated data of third-party drivers. Second, they are also able to steal/read and patch/write the code sections of third-party drivers and Windows core drivers. Finally, they could unlink and modify the allocated structures in Windows internal lists.

To deal with all these three scenarios, I propose to use separate memory access rules to protect dynamically allocated data from being stolen or modified illegally. These rules can also be applied to guarantee the integrity and confidentiality of kernel-mode modules, which are loaded in the memory. The key feature of the memory access rules is that they avoid illegal access to the protected memory without deliberately generating BSOD like Windows built-in security systems (Field, 2006). These rules also allow protecting newly allocated memory regions.

To grant only legal access and prevent all others it is needed to intercept each memory access to the sensitive memory regions. The EPT technology provides an excellent opportunity to trap and process each read-, write-, and execute- access on the 4-kilobyte memory page.

**Scenario 1. Stealing and modifying the allocated data of third-party drivers.** Let us consider the first case when malware tries to access the allocated memory for the third-party driver.

To protect the dynamically allocated data I propose to use the memory access rules, according to which the hypervisor will grant or prevent particular access. The hypervisor controls only the memory regions, whose data is in the list of rules.

This memory access rule includes the following five values: DriverStartAddr, DriverSize, AllocatedStartAddr, AllocatedSize, and SharedAccess. As a result, access attempt to the memory, which is located between AllocatedStartAddr and AllocatedStartAddr + AllocatedSize is granted only to the code from DriverStartAddr to DriverStartAddr + DriverSize. An example of such a rule is given in Table 2.

The default shared access rule prevents read access (R=0) and write access (W=0) to the memory from other drivers and Windows kernel.

If this allocated memory needs to be accessed by another kernel-mode module or Windows kernel I have to add a similar memory access rule. To automatically add a corresponding rule, I use a pre-configured list of driver names, which share the memory with the protected driver, e.g. for sharing allocated memory with Windows Kernel I use ntosknl.exe. Provided I have only a binary code of the driver module, whose allocated memory is critical for stealing and modifying I can apply reverse-engineering analysis to get such a list of driver names.

The list of memory access rules needs to be updated for each kernel-mode module, whose memory is protected. To achieve this, I trap the following events:

- the protected driver is loaded and unloaded;
- the protected module allocates and frees memory.

To realize the aforementioned memory access rules, I leverage the hypervisor facilities and EPT technology using five steps.

**Step 1. Start: trap loading drivers**. First, the hypervisor is loaded before the protected driver is loaded to the memory. The hypervisor will be notified whenever an image is loaded into the memory using PsSetLoadImageNotifyRoutine and choose, which kernel-mode driver will be protected using its name. Apart from module names, the hypervisor can also use the IMAGE_INFO structure content; all these parameters need to be pre-configured.

Table 2 An example of memory access rules

| DriverStartAddr | DriverSize | AllocatedStartAddr | AllocatedSize | SharedAccess |
|---|---|---|---|---|
| fffff8016f670000 (mem_allocator_driver.sys) | 0000B000 | FFFFA400AF3C3F80 | 40 | R=0, W=0 |
| fffff80170201000 (ntosknl.exe) | 008D2000 | FFFFA400AF3C3F80 | 40 | R=0, W=0 |

**Step 2. Trap memory allocation (deallocation) and update EPT pages permissions**. Second, the hypervisor traps each memory allocation (deallocation) routines. The hypervisor will choose only those routines from all that have been called from the code belonging to the protected driver. Without the loss of generality, this paper is essentially concerned with the use of ExAllocatePoolWithTag routine to allocate memory pool and ExFreePoolWithTag to free allocated memory. This function is used in all other memory allocation routines, for example, by ExAllocatePool*, FsRtlAllocatePool*. Also, it is considered that MiAllocatePoolPages routine, which is used by ExAllocatePoolWithTag, has not been called directly by a kernel-mode driver.

The corresponding rules will be added/removed in real time each moment the protected driver calls the allocation/deallocation routine.

By applying EPT technology and EPT paging structures the hypervisor can intercept, process, and control each access to the memory. The proposed algorithm of using EPT facilities is taken from the paper by Korkin & Tanda (2017).

I create EPT paging structures with default page access bits to permit all access. Next using the memory access rules, I limit the access to the fixed data in the kernel mode memory.

After adding a new memory access rule, the hypervisor updates the EPT paging structures: it clears read- and write- permissions on the pages with the protected data and it clears read- and write permissions on the pages with the protected module. After the driver has freed memory, the hypervisor double fills this memory block with zeroes and removes the corresponding memory access rule. Removing the rule will cause restoring the corresponding EPT memory access permissions.

As a result, each read- and write- access to the protected memory will cause an EPT violation.

The hypervisor checks firstly whether an intercepted access belongs to the protected memory. Next, it checks which module has accessed the protected memory, according to the list of memory access rules.

**Step 3. Grant a legitimate access**. The hypervisor grants access to the memory region only for the protected module, which has allocated this memory before, see Table 2. To achieve it the hypervisor temporarily sets read- or write- permission of the protected page and sets a Monitor Trap Flag (MTF). Setting MTF enables the system to generate VM Exit after executing each instruction.

After the legitimate code accesses the memory, the control goes to the hypervisor again because of VM Exit. At this step, the hypervisor restores page permission by clearing access bits and clears MTF.

**Step 4. Prevent an unauthorized access**. If a module not mentioned in the list of rules tries to access the protected memory, the hypervisor needs to prevent it. To achieve it, the hypervisor changes the page frame number (PFN) to the corresponding Page Table Entry (PTE) for the protected memory and temporarily grants access to the replaced memory by setting a read- or write- permission. The hypervisor also sets an MTF.

After an unauthorized module reads or writes to the replaced page and executes just one instruction, the control goes to the hypervisor, because of VM Exit. Next, the hypervisor restores initial configuration: by setting an original PFN value for the protected memory, clearing access bits, and clearing MTF.

**Step 5. Finish: trap unloading the protected driver**. After the protected driver has been unloaded the hypervisor zeroes out the memory, where this driver had been loaded.

To be notified whenever an image is unloaded the hypervisor overwrites the function address of the DriverUnload from the DRIVER_OBJECT (MSDN, n.d.-a; OSR, 2017).

These five steps provide the integrity and confidentiality for the dynamically allocated data in Windows kernel, see Figure 2.

The proposed approach can be used for three malware scenarios, mentioned above.

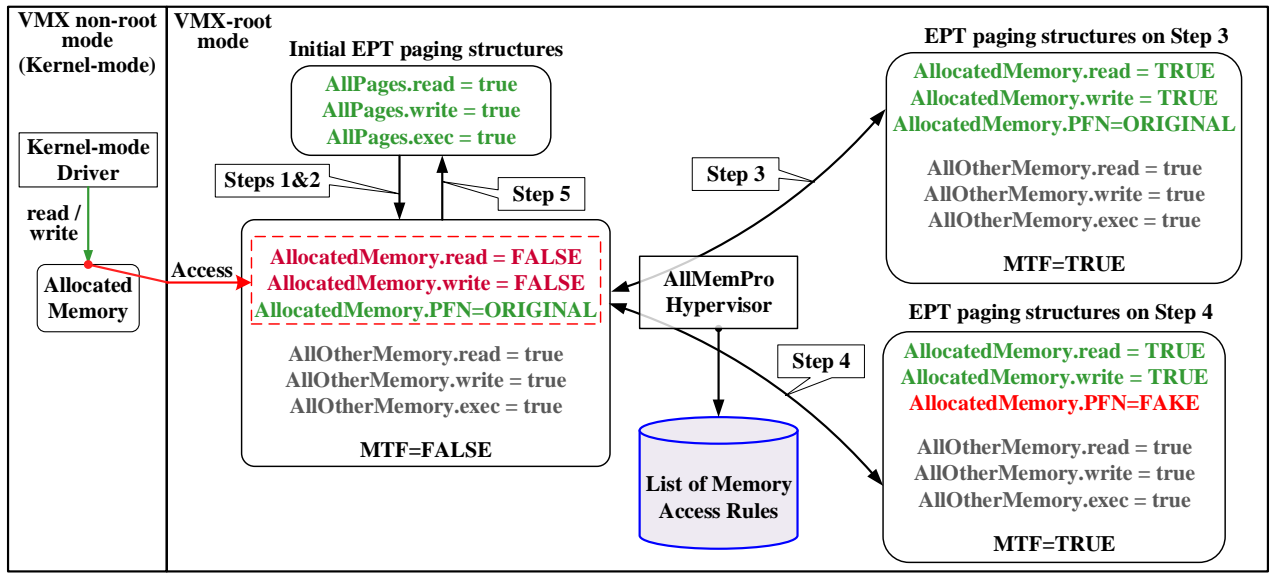

Figure 2 The proposed method of preventing allocated memory from being illegally read or overwritten

**Scenario 2. Stealing and patching code sections.** One of the new protection mechanisms, which has been integrated into Windows Device Guard in Windows 10 is Kernel Mode Code Integrity (KMCI) component. This component prevents modification of executable code directly and does not stop code reading. As a result, the code confidentiality is still becoming vulnerable.

The preliminary testing of AllMemPro shows that the proposed approach of dynamically data protection cannot be used for code protection because it causes a serious overhead. One of the possible ideas for code protection is to apply two EPT structures: first EPT structure allows execution of the protected driver and blocks execution from all other memory; second EPT structure blocks access to the protected data and allows execution from all memory apart from the protected driver.

The code protection will be implemented in further versions of AllMemPro.

**Scenario 3. Tampering Windows Data**. To prevent unlinking and modifying the allocated structures in Windows internal lists I add a memory access rule for each structure (Sherer, 2017).

For example, to avoid hiding EPROCESS structures by DKOM patching with the help of structures unlinking and overwriting its content (Korkin & Nesterow, 2016), the hypervisor adds memory access rules for the existing structures. The hypervisor updates the list of rules by trapping newly allocated structures using PsSetCreateProcessNotifyRoutine routine (MSDN, n.d.-c). In this case, the AllocatedDataAddress and AllocatedDataSize are the address and the size of an EPROCESS structure; the ModuleStartAddress and ModuleSize correspond to the ntoskrnl.exe.

This section has covered a way of maintaining the integrity and confidentiality of dynamically allocated data by using memory access rules and leveraging the hardware-based hypervisor and EPT technology.

The next section covers the architecture of the developed prototype, which realizes memory access rules.

### 3.2. Architecture of AllMemPro

This section covers the design and architecture of the developed hardware-based hypervisor AllMemPro, which protects the dynamically allocated memory in Windows kernel.

The proposed system includes three main components the Controller, Switcher, and Dispatcher.

**The Controller** traps loading drivers and allocation of data. To trap loading of each driver, the Controller uses PsSetLoadImageNotifyRoutine routine, which registers a driver-supplied callback to notify whenever a new driver is loaded. A corresponding callback function gets three basic values, which are used to separate the protected drivers from others: full name to the loaded image file; an ImageBase and an ImageSize of the loaded driver in the memory (MSDN, n.d.-b).

In the current version, the Controller chooses, which driver has to be protected using its name, but it is also possible to choose the protected driver using the calculated CRC from its file in the memory. The Controller intercepts memory allocation routine ExAllocatePoolWithTag and memory deallocation routine ExFreePoolWithTag using DdiMon developed by Tanda (2016). DdiMon monitors and controls kernel API calls with stealth hook using EPT technology.

The Controller intercepts that the protected driver allocates memory and automatically sends the following memory access rule to the Switcher and to the Dispatcher, see Figure 3.

**The Switcher** intercepts access to the protected memory data using the hypervisor and EPT facilities. The code of the Switcher is based on the MemoryMonRWX hypervisor (Korkin & Tanda, 2017).

In the beginning, the Switcher allocates EPT paging structure for all kernel-mode memory pages and sets default access right to skip all read-, write-, and execute- access. After receiving a memory access rule from the Controller, it changes memory access permissions to the pages, which include the protected data by clearing read- and write- bits. As a result, each memory access attempt to the protected data will cause EPT violation.

The Switcher processes all EPT violation and chooses between two possible scenarios: grant and prevent access to the data by calling the Dispatcher.

In the first case, the Switcher allows access to the protected data and sets Monitor Trap Flag (MTF), see EPT structure on step 4 on Figure 2. As a result, after executing just one instruction the Switcher traps control again and restores page permission by clearing read- and write- bits, and clears MTF.

In the second case, as you can see EPT structure on step 4 Figure 2, the Switcher redirects access to the fake page by changing PFN value on the EPT page, which corresponds to the protected data. The Switcher also allows access to this data and sets MTF. As a result, after an unknown code accesses the fake data and executes just one instruction, the control goes to the Switcher again. Now the Switcher restores the original EPT configuration, see steps 1 & 2 in Figure 2.

The Switcher decides which case is processed according to the Dispatcher module.

**The Dispatcher** provides logic to grant and prevent access to the data according to the list of memory access rules.

The Dispatcher grants full privileges to the owner of allocated memory. If an unregistered or unknown code accesses the protected data, the logic of processing will be the following:

- if is_readable==0 a code cannot read the data, otherwise it can read them;
- if 'is_overwritable==0' a code cannot write to this memory, otherwise it can write there;

To allow another driver or Windows Kernel to read or to write to the protected data, the similar memory access rule needs to be added.

In a nutshell, the Dispatcher uses the discretionary access control to prevent illegal access even to one byte of the protected data.

AllMemPro system is developed using Microsoft Visual C++ 2015 with integrated Windows Driver Kit (WDK). It is tested using Vmware Workstation 14 and Windows 10 1709 Build 16299.15 64-bit and multi-core CPU. The source code of AllMemPro is here Korkin (2018).

I can conclude that the proposed hypervisor-based system has the following three advantages:

- it can protect newly allocated memory using the Controller component;
- it can prevent read- and write- access even to one byte of the protected data;
- it works even without the source code of the protected driver.

Next section will cover the three scenarios to demonstrate the facilities of AllMemPro.

```
typedef struct _MEMORY_ACCESS_RULE {
        void*                   drvAddr;
        unsigned __int64        drvSize;
        void*                   allocAddr;
        unsigned __int64        allocSize;
        int                     is_readable;
        int                     is_overwritable;
}MEMORY_ACCESS_RULE, *PMEMORY_ACCESS_RULE;
```

Figure 3 A structure to store a memory access rule

### 3.3. Demos of AllMemPro

This section covers the demonstrations of applying AllMemPro hypervisor to protect kernel-mode memory. I show how AllMemPro isolates the dynamically allocated memory of third-party driver by read- and write- access from another one.

Firstly, I load the kernel-mode driver (mem_allocator_driver.sys), which allocates memory fragment and reads this memory in the loop as well as updates the content of this memory after receiving the IOCTL-code from the console control app (mem_allocator_console.exe).

Next, I load the second kernel-mode driver (mem_attacker_driver.sys), which plays the role of a spyware driver. This driver reads and writes to the content of memory, which was allocated by the first driver. Let me assume that a spyware driver can find the allocated data from mem_allocator_driver.sys without any issues. I control the second driver using another console program.

Figure 4 shows the main scheme. The addresses and sizes of loaded drivers and allocated data are in Table 3.

This unauthorized access demonstrates the fact that the allocated memory is not isolated from unauthorized access from others.

The source code of mem_allocator_driver.sys and mem_attacker_driver.sys with control console apps is here Korkin (2018).

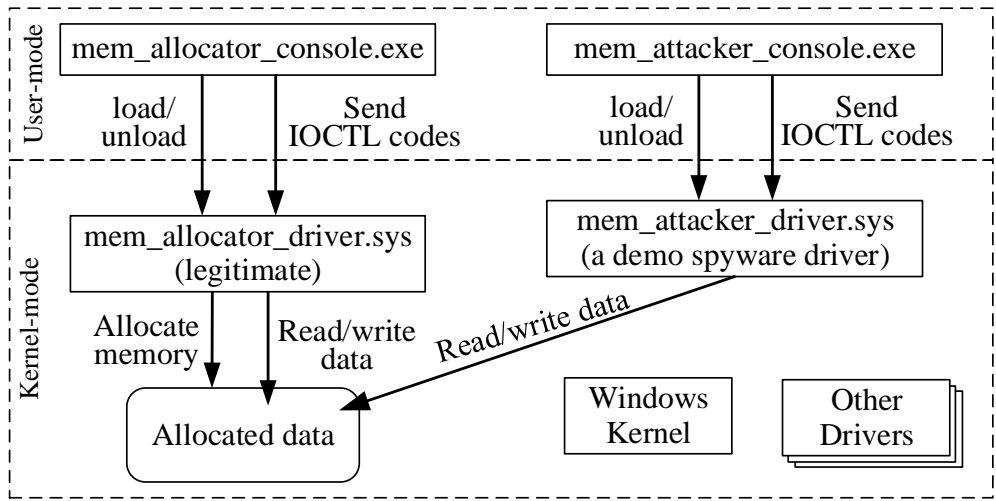

Figure 4 Illegal driver reads and writes the memory allocated by the another driver

Table 3 The details of objects in memory for the Figure 4

| Object in memory | Start address | Size |
|---|---|---|
| mem_allocator_driver.sys | FFFFF8016F630000 | 0xb000 |
| Allocated data by mem_allocator_driver.sys | FFFFA400AC479FD0 | 0x10 |
| mem_attacker_driver.sys.sys | FFFFF8016F650000 | 0x9000 |

Secondly, I load a hypervisor AllMemPro, see Figure 5 and Table 4 and the following memory access rule is added automatically, see Table 5. After that, I restore the allocated data content using control app for the first driver and try to read and write this data using the second driver. I can see that all access attempts from the second driver fail: after reading I get a zero value and writing access does not change the content. The corresponding debug output fragments of AllMemPro are in Figure 6.

I can conclude that AllMemPro provides integrity and confidentiality for the dynamically allocated memory.

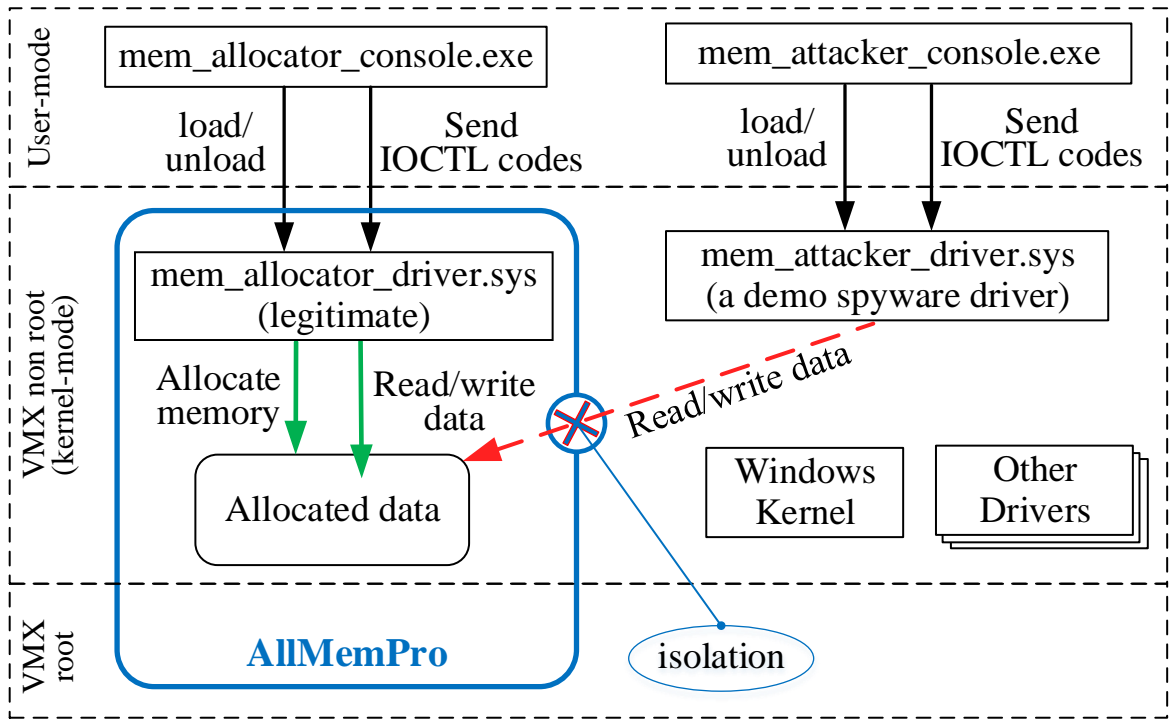

Figure 5 AllMemPro grants access to the allocated memory only to the first kernel-mode driver according to the memory access rule

Table 4 The details of objects in memory for the Figure 5

| Object in memory | Start address | Size |
|---|---|---|
| mem_allocator_driver.sys | FFFFF8016F630000 | 0xb000 |
| Allocated data by mem_allocator_driver.sys | FFFFA400AC479FD0 | 0x10 |
| mem_attacker_driver.sys | FFFFF8016F650000 | 0x9000 |
| nt (ntkrnlmp.exe) | FFFFF80170201000 | 0x8D2000 |

Table 5 The memory access rule allows the mem_allocator_driver.sys access to the allocated data

rule FFFFF8016F630000 B000 FFFFA400AC479FD0 10

**Debug Output Fragment for legal read- access:**
22:34:47.513  INF  #0    4    7732  System
S= FFFFF8016F6317C8 (FFFFF8016F630000), D= FFFFA400AC479FD8 (0000000000000000), T= R

**Debug Output Fragment for legal write- access:**
22:34:50.357  INF  #0    8020  8144  mem_allocator_
S= FFFFF8016F6314EA (FFFFF8016F630000), D= FFFFA400AC479FD8 (0000000000000000), T= W,
00 00 00 00 00 00 00 00 01 0a 00 00 00 00 00 00  => 00 00 00 00 00 00 00 00 ba 0a 00 00 00 00 00 00

**Debug Output Fragment for illegal read-access:**
illegal access FFFFF8016F651228 =>> FFFFA400AC479FD8
 ** RweHandleMonitorTrapFlag FFFFF8016F651228 FFFFA400AC479FD8 **

22:35:05.952  INF  #0    76    8060  mem_allocator_
[Protected via ActiveMemPolice] Memory is being READ. Returning fake contents.
22:35:05.952  INF  #0    76    8060  mem_allocator_
S= FFFFF8016F651228 (FFFFF8016F650000), D= FFFFA400AC479FD8 (0000000000000000), T= R

**Debug Output Fragment: for illegal write-access:**
illegal access FFFFF8016F651257 =>> FFFFA400AC479FD8
 ** RweHandleMonitorTrapFlag FFFFF8016F651257 FFFFA400AC479FD8 **

22:35:20.405  INF  #0    76    8060  mem_allocator_
S= FFFFF8016F651257 (FFFFF8016F650000), D= FFFFA400AC479FD8 (0000000000000000), T= W,
00 00 00 00 00 00 00 00 ba 0a 00 00 00 00 00 00  => 00 00 00 00 00 00 00 00 ba 0a 00 00 00 00 00 00

22:35:20.405  INF  #0    76    8060  mem_allocator_
[Protected via ActiveMemPolice] Memory is being WRITTEN. Returning fake contents.

Figure 6 The fragments of debug output for the Figure 5

Finally, I consider a general case, with shared memory. Now the first driver uses the allocated data to retrieve the system information using Windows Kernel routines, see Figure 7 and Table 6. To share the allocated data between the first driver and Windows Kernel I use the following two memory access rules, see Table 7.

The first line makes the allocated buffer available to the first driver, and the second line – for the Windows kernel, (ntoskrnl.exe). Windows routine has successfully written internal data to this memory. The AllMemPro isolates this data from the second driver. All illegal memory attempts fail, see Figure 8.

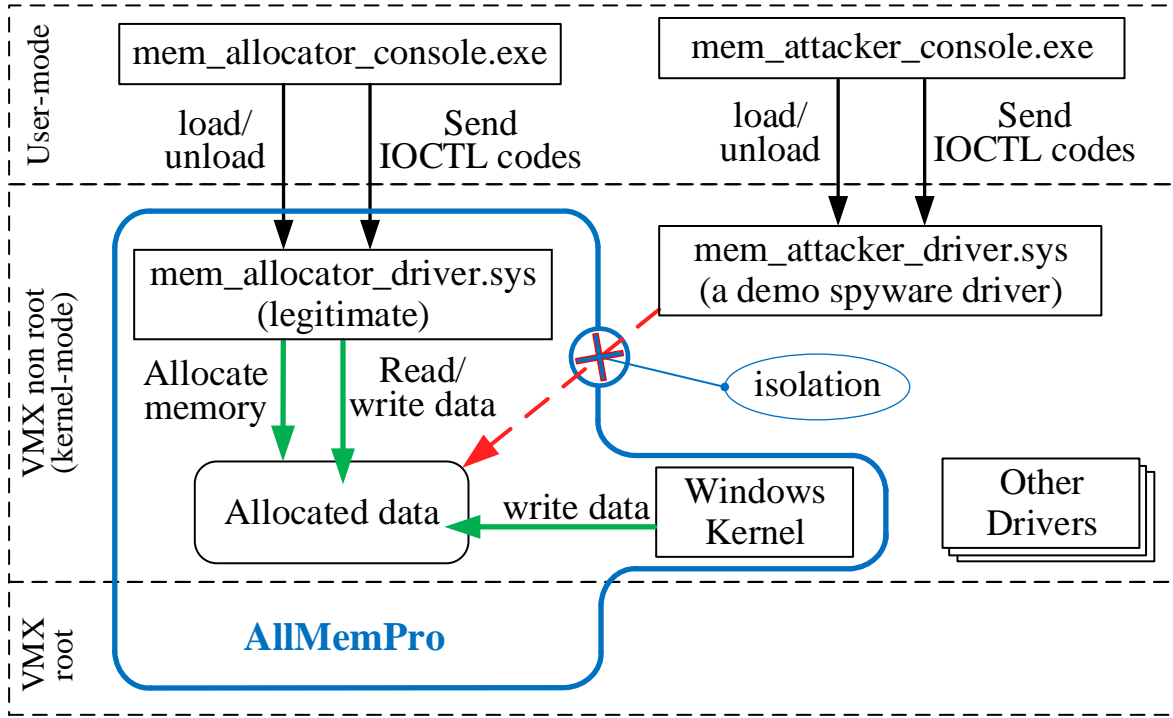

Figure 7 AllMemPro grants access to the allocated memory to the mem_allocator_driver.sys and Windows Kernel. AllMemPro prevents access to the mem_attacker_driver.sys, which is not in the list of rules

Table 6 The details of objects in memory for the Figure 7

| Object in memory | Start address | Size |
|---|---|---|
| mem_allocator_driver.sys | FFFFF8016F630000 | 0xb000 |
| Shared Allocated data | FFFFA400AC479F80 | 0x40 |
| mem_attacker_driver.sys | FFFFF8016F650000 | 0x9000 |
| nt (ntkrnlmp.exe) | FFFFF80170201000 | 0x8D2000 |

Table 7 The rules allow mem_allocator_driver.sys and ntkrnlmp.exe to access to the allocated memory

```
// for mem_allocator_driver.sys
rule FFFFF8016F630000 B000 FFFFA400AC479F80 40

// for ntkrnlmp.exe
rule FFFFF80170201000 8D2000 FFFFA400AC479F80 40
```

**Debug Output Fragment for illegal write- access**
** RweHandleMonitorTrapFlag FFFFF8016F651228 FFFFA400AC479F80 **

22:58:18.560   INF   #0        76      8060   mem_allocator_
[Protected via ActiveMemPolice] Memory is being READ. Returning fake contents.

22:58:18.560   INF   #0        76      8060   mem_allocator_
S= FFFFF8016F651228 (FFFFF8016F650000), D= FFFFA400AC479F80 (0000000000000000), T= R

**Debug Output Fragment for legal write-access (mem_allocator_driver.sys, memset function fragment):**
22:51:03.306   INF   #0        4      7732   System
S= FFFFF8016F631743 (FFFFF8016F630000), D= FFFFA400AC479F8B (0000000000000000), T= W,
00 00 00 00 00 00 00 00 00 00 00 00 7d ff 0f 00  => 00 00 00 00 00 00 00 00 00 00 00 00 7d ff 0f 00

22:51:03.306   INF   #0        4      7732   System
S= FFFFF8016F631743 (FFFFF8016F630000), D= FFFFA400AC479F8C (0000000000000000), T= W,
00 00 00 00 00 00 00 00 00 00 00 00 7d ff 0f 00  => 00 00 00 00 00 00 00 00 00 00 00 00 00 ff 0f 00

22:51:03.306   INF   #0        4      7732   System
S= FFFFF8016F631743 (FFFFF8016F630000), D= FFFFA400AC479F8D (0000000000000000), T= W,
00 00 00 00 00 00 00 00 00 00 00 00 ff 0f 00  => 00 00 00 00 00 00 00 00 00 00 00 00 00 00 0f 00

22:51:03.306   INF   #0        4      7732   System
S= FFFFF8016F631743 (FFFFF8016F630000), D= FFFFA400AC479F8E (0000000000000000), T= W,
00 00 00 00 00 00 00 00 00 00 00 00 0f 00  => 00 00 00 00 00 00 00 00 00 00 00 00 00 00 00 00

22:51:03.306   INF   #0        4      7732   System
S= FFFFF8016F631743 (FFFFF8016F630000), D= FFFFA400AC479F8F (0000000000000000), T= W,
00 00 00 00 00 00 00 00 00 00 00 00 00 00  => 00 00 00 00 00 00 00 00 00 00 00 00 00 00 00 00

**Debug Output Fragment for legal write-access (ntkrnlmp.exe, ZwQuerySystemInformation):**
22:51:03.306   INF   #0        4      7732   System
S= FFFFF801702FB65B (FFFFF80170201000), D= FFFFA400AC479F84 (0000000000000000), T= W,
00 00 00 00 00 00 00 00 00 00 00 00 00 00 00  => 00 00 00 00 5a 62 02 00 00 00 00 00 00 00 00 00

22:51:03.306   INF   #0        4      7732   System
S= FFFFF801702FB65F (FFFFF80170201000), D= FFFFA400AC479F88 (0000000000000000), T= W,
00 00 00 00 5a 62 02 00 00 00 00 00 00 00 00 00  => 00 00 00 00 5a 62 02 00 00 10 00 00 00 00 00 00

22:51:03.306   INF   #0        4      7732   System
S= FFFFF801702FB737 (FFFFF80170201000), D= FFFFA400AC479F8C (0000000000000000), T= W,
00 00 00 00 5a 62 02 00 00 10 00 00 00 00 00  => 00 00 00 00 5a 62 02 00 00 10 00 00 7d ff 0f 00

Figure 8 The fragments of debug output for the Figure 7

In a similar way, I have successfully checked AllMemPro possibility of preventing illegal privileges escalation by directly modifying the content of EPROCESS structure.

As a result, AllMemPro prevents stealing and modifying data, stores in the allocated memory pools in the kernel-mode and moderate performance overhead.

## 4. ALLMEMPRO: POINTS FOR DEVELOPMENT

This chapter focuses on critical analysis of AllMemPro downsides and possible ways of its improvement.

**AllMemPro Overhead**. AllMemPro causes overhead during access to the protected memory regions, and this occurs because of several reasons, which can be partially eliminated.

The evaluation of overhead was performed by measuring the duration of 10 access attempts to the allocated memory in three cases: without hypervisor with enabled memory cache; without hypervisor and disabled cache; and finally, with AllMemPro hypervisor and time cheating.

All these measures are processed on VMware Workstation Pro in the release version of all drivers. To get enough measurements, I use 200 repetitions, next I delete five maximums values and five minimums ones and finally calculate the average and deviation values, see Table 8.

In the first case, the latency is quite small because after first several memory access attempts the corresponding virtual and physical addresses are cached and further access was processed using these cache values. To make the comparison with hypervisor case a bit more appropriate I applied the second case with the deliberately disabled cache.

In the third case, I measured the latency of memory access to the protected memory from the legal driver, when AllMemPro has been activated.

AllMemPro hypervisor traps each access to the protected memory region because this memory does not have read- and write- permissions. After that, the hypervisor sets the corresponding permissions and according to the memory access rule allows or disallows memory access to this data by changing PFN-value. At this step, AllMemPro sets MTF and returns control to the guest. After the guest executes just one instruction, the control goes to the hypervisor again because of MTF.

Next AllMemPro clears MTF and restores original permissions to the memory to be able to trap a new access to the protected region.

As a result, for each memory access attempt, AllMemPro has been called two times, which leads to time overhead. It is possible to reduce it by applying two EPT structures. The first EPT structure corresponds to the legal driver and its memory and the second EPT – to the other memory ranges. However, in this case, to isolate the allocated memory of two and more drivers from each other and from the other drivers it is needed to allocate the separate EPT structures for each driver. This approach has been implemented in the MemoryMonRWX hypervisor by Korkin and Tanda (2017).

As a result, AllMemPro protects memory, which has a low frequency of access, for example, the EPROCESS.Token value. AllMemPro does not decrease memory access time for non-protected memory regions. To protect memory, which is very often accessible it is possible to apply multiple EPT structures.

Table 8 Time evaluation

| No. | Cases | Memory Access Latency, TSC ticks |
|---|---|---|
| 1 | without AllMemPro with enabled cache | 70±2 |
| 2 | without AllMemPro with disabled cache | 100.000±4.000 |
| 3 | with AllMemPro hypervisor | 500.000±10.000 |

**Indirect Memory Access**. AllMemPro determines the source address by reading the value of RIP register from VMCS-structure. To prevent indirect memory access AllMemPro can additionally check the call stack, but in the current version, this is not implemented.

**Self-protection: Resilience to Manipulations**. Malware driver can access the protected data by deliberately changing the memory access rules content. It is possible to protect these rules by applying the proposed AllMemPro hypervisor to protect itself.

**Protect Memory with Shared Access**. The current AllMemPro protects shared memory in the following way. For two drivers, using shared memory, all memory regions, which are allocated to each of these drivers, are available to read- and write- access by either driver and are isolated from any other access. It is possible to provide fine-grained access control to shared memory.

To allow shared access only to the programmer-specified buffers, it is necessary to integrate AllMemPro at a source level during driver development.

**Pagefile Mechanism**. It is possible to overwrite the third-party driver allocated memory by forcing the kernel to page-out the kernel memory pool and then locating and overwriting the driver memory inside the pagefile in the hard disk. This can be used not only to attack the memory pool but also to overwrite the third-party driver code sections. The current version of AllMemPro does not protect the pagefile mechanism.

**Firmware exploitation as a vector of infection**. This research does not consider firmware exploitation as one of the possible ways of infections. Because of this infection, the malware code is able to tamper both OS and hypervisor memory, as well as injecting code into OS kernel. Hypervisor-based solutions cannot prevent such infections.

**Direct Access to the Physical Memory**. AllMemPro can potentially prevent direct access to the physical memory or access to the mapped memory pages by MmMapLockedPages(). The current version of AllMemPro deliberately converts the virtual address to the physical one. At the same time, DMA attacks using firmware exploitation and hardware are out of the scope of this paper.

**Confidentiality and Integrity of Code Sections**. The current version of AllMemPro protects only allocated data in the kernel mode. The protection of code sections from being illegally read and overwritten will be implemented further.

**Joint work with Win10 Windows 10: Device Guard and Credential Guard.** AllMemPro has been successfully tested on default installation Windows 10 x64 1709 version, which is installed as a BIOS-version. The tested on UEFI versions of Windows OS will be processed further.

**SGX technology and Virtual Secure Mode**. The Software Guard Extensions (SGX) technology makes it possible to protect the areas of execution in memory via enclaves. This technology has been integrated into 6<sup>th</sup> generation Intel CPUs, while AllMemPro supports Intel CPUs since Nehalem microarchitecture, which is more common now.

In addition, a similar idea was implemented to secure kernel for Windows 10 by leveraging Virtual Secure Mode with Virtual Trust Levels (VTLs). According to A. Ionescu, it is possible to apply VTL to protect some kernel-mode data (Juarez, 2015; Ionescu, 2015; Laiho, 2016).

## 5. CONCLUSIONS & FUTURE WORK

To sum up, the proposed security system AllMemPro has the following competitive advantages:

- it provides fine-grained control to mediate access from kernel-mode drivers to the dynamically allocated memory;
- it protects allocated memory of third-party drivers and the content of OS structures;
- it guarantees the integrity and confidentiality of the allocated data by redirecting unauthorized access without crushing OS;
- it is an open-source project with minimal lines of code, which can be used for educational purposes to teach VT-x & EPT.

**Spectre & Meltdown Attacks**. AllMemPro hypervisor seems to be able to prevent sensitive kernel-mode data from being stolen using the newest Spectre and Meltdown attacks (Horn, 2018). However, further research is required.

With regard to the future, I would like to suggest the following ideas of using AllMemPro to prevent:

- o leakage of the Windows Telemetry memory data;
- o drivers' exploitation by validating kernel-mode code execution;
- o unauthorized access from kernel-mode malware to files, registry, and processes.

**Windows Telemetry leakage**. Windows Telemetry includes a lot of sensitive user information and has to be protected from unauthorized access by malware. Another issue is to disable the Windows Telemetry data reliably. As a result, users will be confident that the private data: Browsing History, Voice Input, GPS data, etc. will not be collected and transferred to anyone.

**Preventing Drivers' Exploitation**. To reveal the fact that the driver is being exploited I propose the following. AllMemPro will trap and log the driver code execution using a lot of valid input data. Secondly, I will analyze the progress of code execution and create some signatures, using the corresponding control flow graph (CFG). Finally, I will test this code using common data or data with exploits. By comparing the code execution with signature CFG I will check whether the code executes all its parts or it skips something from CFG. If it skips any part, it means that the driver's behavior is not normal and someone is using its vulnerability.

**Preventing Kernel-Mode Malware from Accessing Files and Registry.** Windows security model provides the registry key and file security only for user-mode applications. It means that kernel-mode drivers do not have any limitations to access filesystem and registry. As a result malware driver can read, write, and even delete files and registry data, which are processed by user-mode applications or other drivers. My idea is to adapt the AllMemPro to prevent this unauthorized access by monitoring and controlling access attempts to filesystem and registry. The proposed system will use the similar access rules to grant access only to the owner and registered drivers and will stop access from the illegal ones.

## 6. ACKNOWLEDGMENTS

I thank the anonymous reviewers for their constructive feedback on this work.

I would like to thank Michael Chaney, alumnus of Embry Riddle Aeronautical University, Daytona Beach, Florida, US, Ashlyn King, an intern at Russian Flagship Center, University of Wisconsin Madison, Madison, Wisconsin, US, and Veronika Domova, a researcher at ABB and Linkoping University, Sweden for their time and effort in proofreading this paper.

I would like to especially thank Eugene Rodionov, a senior security researcher at Intel for his thorough analysis of this paper and reasonable comments, which significantly improved the research.

In addition, I would like to thank the following security experts for their valuable comments: Elia Florio, Bruce Dang, Nicola Cowie, Alex Ionescu, Edgar Barbosa, Mariano Graziano, Oguzhan Filizlibay, David Kapman, Alex Tereshkin, Vitaliy Malkin, Nicolas Alejandro Economou, Jonathan Brossard, Artem Baranov, Mohamed Saher, Alexandre Borges, and Kelvin Chan.